\documentclass[10pt]{article}
\usepackage{latexsym}
\usepackage{amssymb}
\usepackage{amsmath}
\usepackage{amscd}
\usepackage{amsthm}
\usepackage[left=2cm,top=2.5cm,right=2.5cm,bottom=1.5cm]{geometry}
\usepackage[dvips]{graphicx}
\usepackage{hyperref}
\begin{document}
\begin{center}
\large{\bf{Two-fluid scenario for dark energy models in an FRW universe-revisited}} \\
\vspace{10mm}
\normalsize{Bijan Saha$^1$, Hassan Amirhashchi$^2$, Anirudh Pradhan$^3$}\\
\vspace{5mm}
\normalsize{$^{1}$Laboratory of Information Technologies, Joint Institute for Nuclear Research,
141980 Dubna, Russia \\
$^3$E-mail: bijan@jinr.ru} \\
\vspace{5mm}
\normalsize{$^{2}$Young Researchers Club, Mahshahr Branch, Islamic Azad University, Mahshahr, Iran \\
E-mail: h.amirhashchi@mahshahriau.ac.ir} \\
\vspace{5mm}
\normalsize{$^{3}$Department of Mathematics, Hindu Post-graduate College, Zamania-232 331,
Ghazipur, India \\
E-mail: pradhan.anirudh@gmail.com} \\
\end{center}
\vspace{10mm}
\begin{abstract}
In this paper we study the evolution of the dark energy parameter within the scope of a spatially homogeneous 
and isotropic Friedmann-Robertson-Walker (FRW) model filled with barotropic fluid and dark energy by revisiting 
the recent results (Amirhashchi et al. in Chin. Phys. Lett. 28:039801, 2011a). To prevail the deterministic solution 
we select the scale factor $a(t) = \sqrt{t^{n}e^{t}}$ which generates a time-dependent deceleration parameter (DP), 
representing a model which generates a transition of the universe from the early decelerating phase to the recent 
accelerating phase. We consider the two cases of an interacting and non-interacting two-fluid (barotropic and dark energy) 
scenario and obtained general results. The cosmic jerk parameter in our derived model is also found to be in good agreement 
with the recent data of astrophysical observations under the suitable condition. The physical aspects of the models and 
the stability of the corresponding solutions are also discussed.   
\end{abstract}

\section{Introduction}
The use of Type Ia supernovae as standardized light sources $-$ {\bf calibrated candles} $-$ led to the observational 
discovery of dark energy by two groups in 1998 (Riess et al. 1998, Perlmutter et al. 1999). Before the accelerated 
expansion of the universe was revealed by high red-shift supernovae Ia (SNe Ia) observations (Riess et al. 1998, 
Perlmutter et al. 1999), it could hardly be presumed that the main ingredients of the universe are dark sectors. 
The concept of dark energy was proposed for understanding this currently accelerating expansion of the universe, 
and then its existence was confirmed by several high precision observational experiments (Bennett et al. 2003; 
Spergel et al. 2003; Tegmark et al. 2004; Abazajian et al. 2004; Hawkins et al. 2003; Verde et al. 2002), especially 
the Wilkinson Microwave Anisotropy Probe (WMAP) satellite experiment. The WMAP shows that dark energy occupies about 
$73\%$of the energy of the universe, and dark matter about $23\%$. The usual baryon matter, which can be described by 
our known particle theory, occupies only about $4\%$ of the total energy of the universe. Measurements as of 2008, with 
the greatest weight coming from the combination of supernovae with either cosmic microwave background or baryon acoustic 
oscillation data, show that dark energy makes up $72\pm3\%$ of the total energy density of the universe, and its 
equation of state averaged over the last 7 billion years is $\omega = −1.00 \pm 0.1$ (Kowalski et al. 2008). This is 
consistent with the simplest picture, the cosmological constant, but also with a great many scenarios of time varying 
dark energy or extended gravity theories. In order to explain why the cosmic acceleration happens, many theories have 
been proposed. Although theories of trying to modify Einstein equations constitute a big part of these attempt, the 
mainstream explanation for this problem, however, is known as theories of dark energies. It is the most accepted idea 
that a mysterious dominant component, dark energy (DE), with negative pressure, leads to this cosmic acceleration, though 
its nature and cosmological origin still remain enigmatic at present. In the concordance model, the energy content of the 
Universe is dominated by a cosmological constant $\Lambda \simeq 1.7\times 10^{-66} (eV)^{2}$ such that $\Omega_{\Lambda} 
= \Lambda/(3H^{2}_{0}) \simeq 0.73$. Here $H_{0}$ denotes the Hubble constant which we parameterize as $H_{0} = 100 ~ hkm 
~ s^{-1} ~ Mpc^{-1} = 2.1332 ~h\times10^{-33} eV$. The second component of the concordance model is pressure-less matter
with $\Omega_{m} = \rho_{m}/\rho_{c} = \rho_{m}/(3H^{2}_{0}/8\pi G) \simeq 0.13/h^{2}$, where G is Newton's constant.\\

The simplest candidate of dark energy is the cosmological constant. It is, however, plagued with the so-called 
coincidence problem and the cosmological constant problem (Weinberg 1989; Sahni and Starobinsky 2000; 
Peebles and Ratra 2003; Padmanabhan 2003; Copeland et al. 2006). Another possible form of dark energy is provided 
by scalar fields. Thus some dynamical scalar field, such as quintessence (Wetterich 1988; Ratra and Peebles 1988; 
Caldwell et al. 1998), phantom (Caldwell 2002; Nojiri and Odintsov 2003, 2005; Sahni and Shtanov 2003; Xiangyun et 
al. 2008), quintom (Elizalde et al. 2004; Feng et al. 2005; Setare 2006; Cai et al. 2007) and K-essence (Alimohammadi 
and Sadjadi 2006; Wei et al. 2007), are proposed as possible candidate of dark energy. However, it is worth mentioned 
here that for these scalar field models the coincident problem still remains. Although the two dark components are usually 
studied under the assumption that there is no interaction between them, one cannot exclude such a possibility. In fact, 
researches show that a presumed interaction may help alleviate the coincident problem (Chimento et al. 2003; Chimento 
and Pavon 2006). A more comprehensive review is provided in Copeland et al. (2006). \\

If $\rho_{D}$ and $p_{D}$ are density and pressure, respectively, of the DE can be characterized by the equation-of-state 
(EoE) parameter $\omega_{D}$, defined by $\omega_{D} = \frac{p_{D}}{\rho_{D}}$ which is negative for DE. According to the 
latest cosmological data available, the uncertainties are still too large to discriminate among the three cases $\omega < −1$, 
$\omega = −1$, and $\omega > −1$: $\omega = −1.04^{+0.09}_{-0.10}$ (Amanullah et al. 2010). Since the quintessence has the 
property with the EoS $\omega > -1$ and the phantom behaves as $\omega < -1$, we can speculate that the quintom dark energy 
is a two-component system containing quintessence and phantom. The of DE increases with the increase of scale factor, and 
both the scale factor and the phantom energy density can become infinite at a finite time $t$, a condition known as the 
``big rip'' (Caldwell 2002; Caldwell et al. 2003; Frampton and Takahashi 2003; Nesseris and Perivolaropoulos 2004). 
The closer examination shows that the condition $\omega < -1$ is not sufficient for a singularity occurrence. First of 
all, a transition phantom cosmology is quite possible. Recently, a new scenario to avoid a future singularity has been 
proposed by Frampton et al. (2011, 2012a, 2012b). In this scenario, the EoS parameter $\omega < -1$, so the dark energy 
density increases with time, but $\omega$ approaches $-1$ asymptotically and sufficiently rapidly that a singularity is 
avoided. This proposed non-singular cosmology was termed as a ``little rip'' (Brevik et al. 2011; Nojiri et al. 2012; Granda 
and Loaiza 2012; Xi et al. 2012; Astashenok et al. 2012 and references therein). The evolution of the little rip cosmology 
is close to that of $\Lambda$CDM up to the present, and is similarly consistent with the observational data. \\

The cosmological evolution of a two-field dilation model of dark energy was investigated by Liang et al. (2009). 
The viscous dark tachyon cosmology in interacting and non-interacting cases in non-flat FRW Universe was studied by Setare 
et al. (2009). Recently, Amirhashchi et al. (2011a, 2011b) and Pradhan et al. (2011a) have studied an interacting and 
non-interacting two-fluid scenario for dark energy models in FRW universe. In this report we study the evolution of the 
dark energy parameter within the framework of a FRW cosmological model filled with two fluids (barotropic and dark energy) 
by revisiting the recent work of Amirhashchi et al. (2011a) and obtained more general results. The cosmological implications 
of this two-fluid scenario will be discussed in detail in this paper. In doing so we consider both non-interacting and 
interacting cases. The out line of the paper is as follows: In Sect. $2$, the metric and the basic equations are described. 
Sections $3$ and $4$ deal with non-interacting and interacting two-fluid models respectively and their physical significances. 
Physical acceptability and the stability of corresponding solutions are analyzed in Sect. $5$. Finally, conclusions are 
summarized in the last Sect. $6$.  
\section{The metric and basic equations}
We consider the spherically symmetric Friedmann-Robertson-Walker (FRW) metric as
\begin{equation}
\label{eq1}
ds^{2} = -dt^{2} + a^{2}(t)\left[\frac{dr^{2}}{1 - kr^{2}} + r^{2}d\Omega^{2}\right],
\end{equation}
here $a(t)$ is the scale factor and the curvature constant $k$ is $-1, 0, +1$ respectively
for open, flat and closed models
of the universe.\\

The Einstein's field equations (with $8\pi G = 1$ and $c = 1$) read as
\begin{equation}
\label{eq2}
R^{j}_{i} - \frac{1}{2}R\delta^{j}_{i} =  T^{j}_{i},
\end{equation}
where the symbols have their usual meaning and $T^{j}_{i}$ is the two-fluid energy-momentum tensor consisting of
dark energy and barotropic fluid.\\

In a co-moving coordinate system, Einstein's field equations (\ref{eq2}) for the line element (\ref{eq1}) lead to
\begin{equation}
\label{eq3} p_{tot} = -\left(2\frac{\ddot{a}}{a} + \frac{\dot{a}^{2}}{a^{2}} + \frac{k}{a^{2}}\right),
\end{equation}
and
\begin{equation}
\label{eq4}\rho_{tot} = 3\left(\frac{\dot{a}^{2}}{a^{2}} + \frac{k}{a^{2}}\right),
\end{equation}
where $p_{tot} = p_{m} + p_{D}$ and $\rho_{tot} = \rho_{m} + \rho_{D}$. Here $p_{m}$ and $\rho_{m}$ are pressure
and energy density of barotropic fluid and $p_{D}$ \& $\rho_{D}$ are pressure and energy density of dark fluid
respectively.\\

The Bianchi identity $G^{;j}_{ij} = 0$ leads to  $T^{;j}_{ij} = 0$ which yields
\begin{equation}
\label{eq5}\dot{\rho}_{tot} + 3\frac{\dot{a}}{a}(\rho_{tot} + p_{tot})=0.
\end{equation}
The EoS of the barotropic fluid  and dark field are given by
\begin{equation}
\label{eq6}\omega_{m} = \frac{p_{m}}{\rho_{m}},
\end{equation}
and
\begin{equation}
\label{eq7}\omega_{D} = \frac{p_{D}}{\rho_{D}},
\end{equation}
respectively. In the following sections we deal with two cases, (i) non-interacting two-fluid model and (ii)
interacting two-fluid model. 
\begin{figure}[htbp]
\centering
\includegraphics[width=8cm,height=8cm,angle=0]{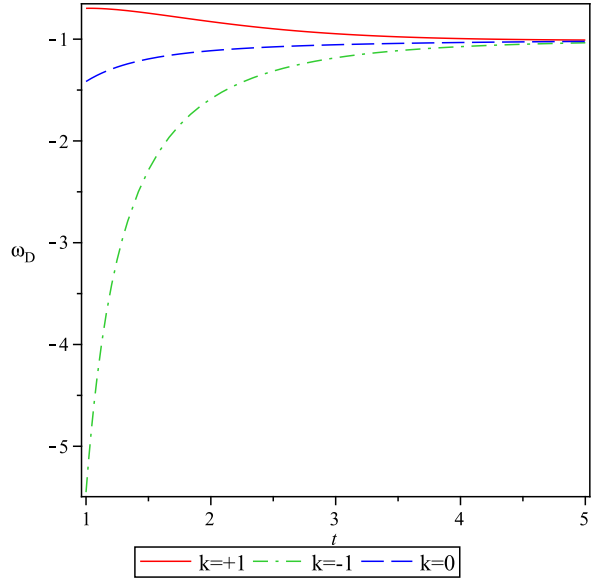}
\caption{The plot of EoS parameter ($\omega_{D}$) Vs. $t$ for non-interacting two-fluid scenario. 
Here $\rho_{0}=1$, $\omega_{m}=0.5$, $n=\frac{1}{2}$}
\end{figure}
\section{Non-interacting two-fluid model}
First, we consider that two fluids do not interact with each other. Therefor, the general form of conservation
Eq. (\ref{eq5}) leads us to write the conservation equation for the dark and barotropic fluid separately as,
\begin{equation}
\label{eq8}\dot{\rho}_{m} + 3\frac{\dot{a}}{a}\left(\rho_{m} + p_{m}\right) = 0,
\end{equation}
and
\begin{equation}
\label{eq9}\dot{\rho}_{D} + 3\frac{\dot{a}}{a}\left(\rho_{D} + p_{D}\right) = 0.
\end{equation}
Here is, of course, a structural difference between Eqs. (8) and (9). Because Eq. (8) is in the form of 
$\omega_{m}$ which is constant and hence Eq. (8) is integrable. But Eq. (9) is a function of $\omega_{D}$. 
Accordingly, $\rho_{D}$ and $p_{D}$ are also function of $\omega_{D}$. Therefore, we can not integrate 
Eq. (9) as it is a function of $\omega_{D}$ which is an unknown time dependent parameter. Integration of 
Eq. (\ref{eq8}) leads to
\begin{equation}
\label{eq10}\rho_{m} = \rho_{0}a^{-3(1 + \omega_{m})}.
\end{equation}
By using Eq. (\ref{eq10}) in Eqs. (\ref{eq3}) and (\ref{eq4}), we first obtain the $\rho_{D}$ and $p_{D}$ in 
term of scale factor $a(t)$
\begin{equation}
\label{eq11}\rho_{D} = 3\left(\frac{\dot{a}^{2}}{a^{2}} + \frac{k}{a^{2}}\right) - \rho_{0}a^{-3(1 + \omega_{m})}.
\end{equation}
and
\begin{equation}
\label{eq12} p_{D} = -\left(2\frac{\ddot{a}}{a} + \frac{\dot{a}^{2}}{a^{2}} + \frac{k}{a^{2}}\right) - \rho_{0}\omega_{m}
a^{-3(1 + \omega_{m})}.
\end{equation}
In literature it is common to use a constant deceleration parameter (Akarsu and Kilinc 2010a, 2010b; Amirhashchi et al. 2011c; 
Pradhan et al. 2011a, 2011b; Kumar and Yadav 2011; Yadav 2011; Kumar and Singh 2011), as it duly gives a power law for metric 
function or corresponding quantity. The motivation to choose such time dependent DP is behind the fact that the universe is 
accelerated expansion at present as observed in recent observations of Type Ia supernova (Riess et al. 1998, 2004; Perlmutter et al. 
1999; Tonry et al. 2003; Clocchiatti et al. 2006) and CMB anisotropies (Bennett et al. 2003; de Bernardis et al. 1998; Hanany et al. 
2000) and decelerated expansion in the past. Also, the transition redshift from deceleration expansion to accelerated expansion is 
about 0.5. Now for a Universe which was decelerating in past and accelerating at the present time, the DP must show signature 
flipping (see the Refs. Padmanabhan and Roychowdhury 2003; Amendola 2003; Riess et al. 2001). So, in general, the DP is not 
a constant but time variable. The motivation to choose the following scale factor is that it provides a time-dependent DP.\\

Under above motivations we take following {\it ansatz} for the scale factor, where increase in term of time evolution is
\begin{equation}
\label{eq13} a(t) = \sqrt{t^{n}e^{t}},
\end{equation}
where $n$ is a positive constant. If we put $n = 0$, Eq. (\ref{eq13}) reduces to $a(t) = \sqrt{e^{t}}$ i.e. exponential law 
of variation. This choice of scale factor yields a non-singular cosmology called a ``little rip`` which 
is discussed below. It is worth mention here that the solutions in both non-interacting and interacting models do not blow up 
at any given epoch for the choice of this {\it ansatz}. It should be remembered that $a(t)$ is a unit less function. 
In Eq. (\ref{eq13}), $a$ is a function of $\tau = \frac{t}{t_{1}}$, where $t_{1}$ is a constant of unit [Time]. As a result, 
being a function time, $a$ still remains unit less. For simplicity here and further we write $a$ as a function of $t$ with $t$ 
now being unit less. This {\it ansatz} generalizes the one proposed in (Amirhashchi et al. 2011a; Pradhan et al. 2012). Recently, 
the {\it ansatz} (\ref{eq13}) is also used by Pradhan \& Amirhashchi (2011a) in studying the accelerating DE models in Bianchi 
type-V space-time. \\

\begin{figure}[htbp]
\centering
\includegraphics[width=8cm,height=8cm,angle=0]{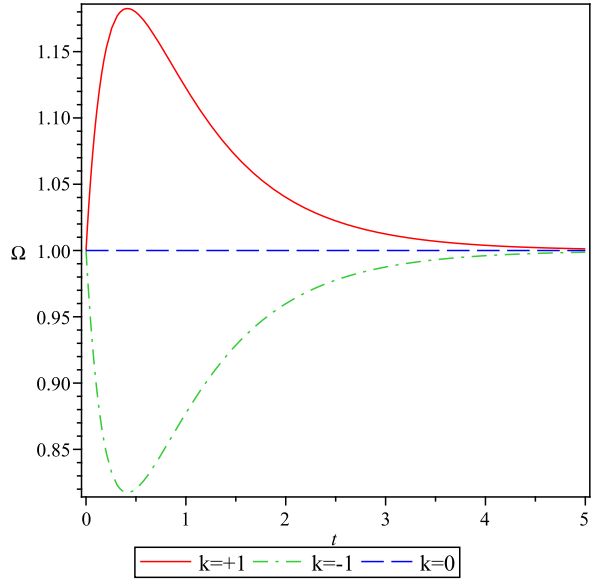}
\caption{The plot of density parameter ($\Omega$) Vs. $t$ for $n=\frac{1}{2}$}
\end{figure}
\begin{figure}[htbp]
\centering
\includegraphics[width=8cm,height=8cm,angle=0]{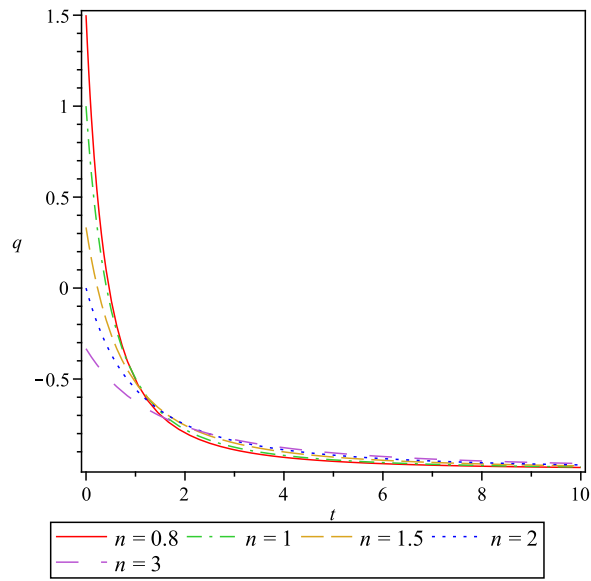}
\caption{The plot of deceleration parameter ($q$) Vs. $t$}
\end{figure}
We define the deceleration parameter $q$ as usual, i.e.
\begin{equation}
\label{eq14}q = - \frac{\ddot{a} a}{\dot{a}^{2}} = - \frac{\ddot{a}}{aH^{2}}.
\end{equation}
Using Eq. (\ref{eq13}) into Eq. (\ref{eq20}), we find
\begin{equation}
\label{eq15}q = \frac{2n}{(n+t)^{2}}-1.
\end{equation}
From Eq. (\ref{eq15}), we observe that $q > 0$ for $t < \sqrt{2n} - n$ and $q < 0$ for $t > \sqrt{2n} -n$. 
It is observed that for $0 < n < 2$, our model is evolving from deceleration phase to acceleration phase. Also, recent 
observations of SNe Ia, expose that the present universe is accelerating and the value of DP lies to some place in the 
range $-1 < q < 0$. It follows that in our derived model, one can choose the value of DP consistent with the observation. 
Figure $3$ depicts the deceleration parameter ($q$) versus time which gives the behaviour of $q$ from decelerating to 
accelerating phase for different values of $n$. \\

By using this scale factor in Eqs. (\ref{eq11}) and (\ref{eq12}), the $\rho_{D}$ and  $p_{D}$ are obtained as
\begin{equation}
\label{eq16} \rho_{D} = 3\left(\frac{n + t}{2t}\right)^{2}+\frac{3k}{t^{n}e^{t}} - \rho_{0}(t^{n}e^{t})^
{-\frac{3}{2}(1 + \omega_{m})},
\end{equation}
and
\begin{equation}
\label{eq17} p_{D} = -\left[3\left(\frac{n + t}{2t}\right)^{2} - \frac{n}{t^{2}} + \frac{k}{t^{n}e^{t}} - 
\rho_{0}\omega_{m}(t^{n}e^{t})^{-\frac{3}{2}(1 + \omega_{m})}\right],
\end{equation}
respectively. By using Eqs. (\ref{eq16}) and (\ref{eq17}) in Eq. (\ref{eq7}), we can find the equation of 
state of dark field in term of time as
\begin{equation}
\label{eq18}\omega_{D} = -\frac{3\left(\frac{n + t}{2t}\right)^{2} - \frac{n}{t^{2}} + \frac{k}{t^{n}e^{t}} 
- \rho_{0}\omega_{m}(t^{n}e^{t})^{-\frac{3}{2}(1 + \omega_{m})}}{\left(\frac{n + t}{2t}\right)^{2} + 
\frac{k}{t^{n}e^{t}} - \rho_{0}(t^{n}e^{t})^{-\frac{3}{2}(1 + \omega_{m})}}.
\end{equation}
The behavior of EoS for DE ($\omega_{D}$) in term of cosmic time $t$ is shown in Fig. $1$. It is observed that 
for closed universe the $\omega_{D}$ is decreasing function of time whereas for open and flat universes the EoS 
parameter is an increasing function of time, the rapidity of their growth at the early stage depends on the type 
of the universes, while later on they all tend to the same constant value independent to it. From this figure we also 
observe that the EoS parameters of closed, open and flat universes are varying in quintessence ($\omega_{D} > -1$), 
phantom ($-3 < \omega_{D} < -1$) and Super phantom ($\omega_{D} < -3$) regions respectively, while later on they 
tend to the same constant value $-1$ (i.e. cosmological constant) independent to it. \\

From Fig. $1$, it is observed that in the case of open and flat universes $\omega_{D}$ is less than $-1$, so the DE 
density increases with time, but $\omega_{D}$ approaches $-1$ asymptotically and sufficiently rapidly that a singularity 
is avoided. This is the scenario of a ``little rip'' Frampton et al. (2011, 2012a, 2012b).  \\ 

\begin{figure}[htbp]
\centering
\includegraphics[width=8cm,height=8cm,angle=0]{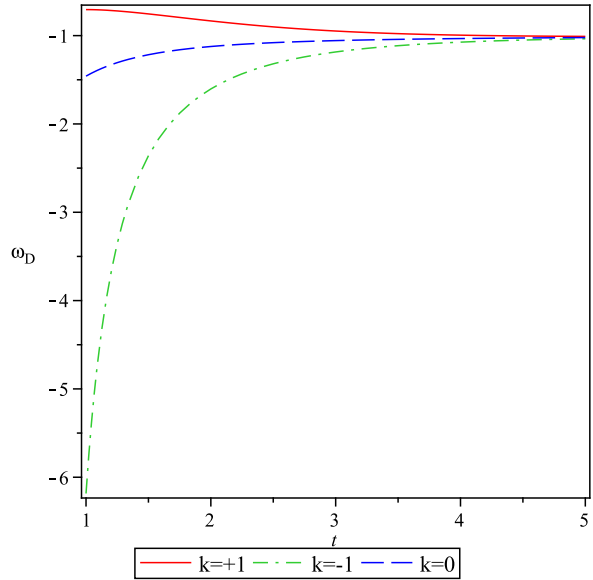}
\caption{The plot of EoS parameter $\omega_{d}$ Vs. $t$ for interacting two-fluid scenario. 
Here $\rho_{0} = 1$, $\omega_{m} = 0.5$, $n = \frac{1}{2}$, $\sigma = 0.3 $}
\end{figure}
\begin{figure}[htbp]
\centering
\includegraphics[width=8cm,height=8cm,angle=0]{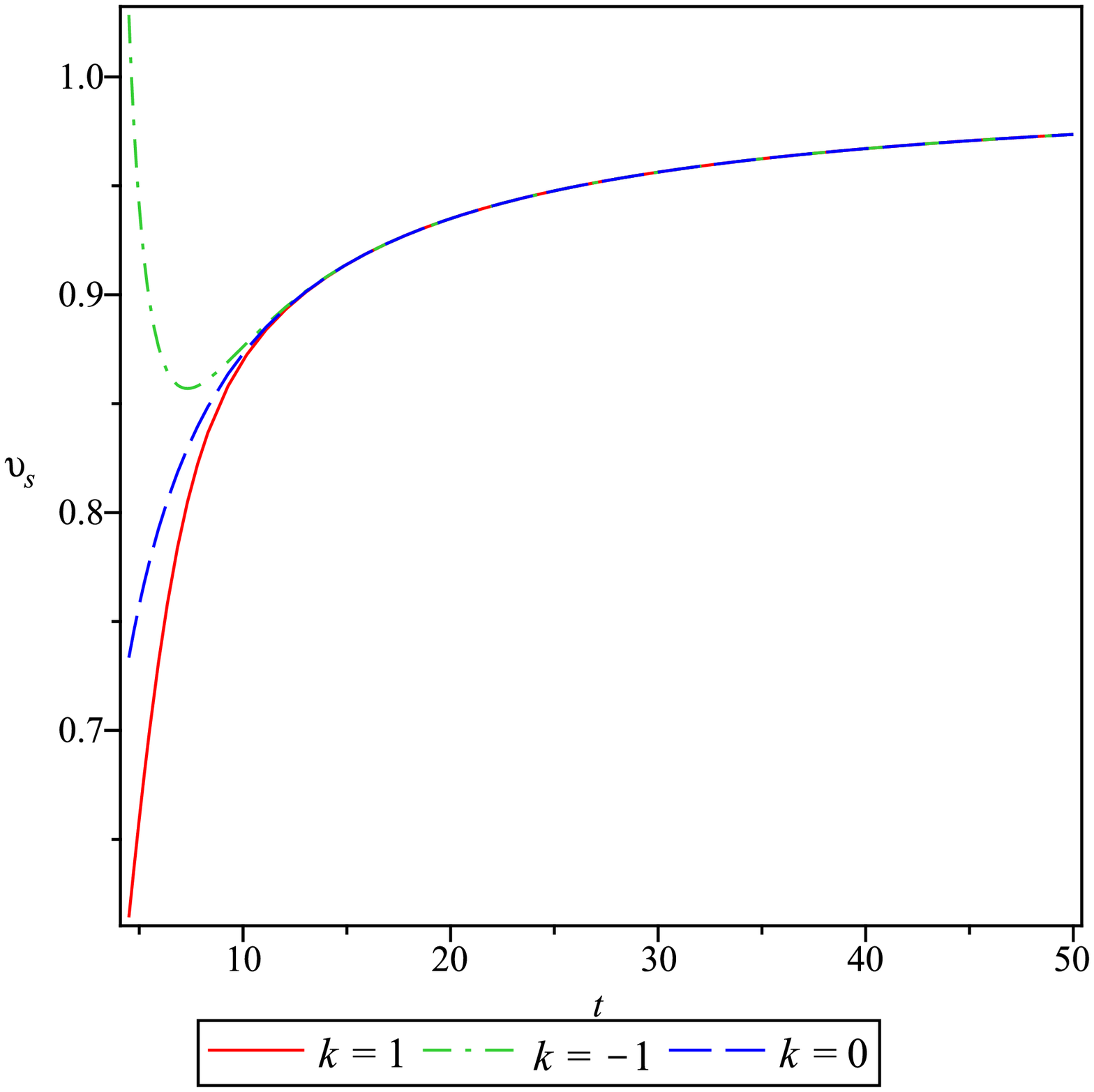}
\caption{The plot of sound speed $\upsilon_{s} $ Vs. $t$ for non-interacting two-fluid scenario}
\end{figure}
\begin{figure}[htbp]
\centering
\includegraphics[width=8cm,height=8cm,angle=0]{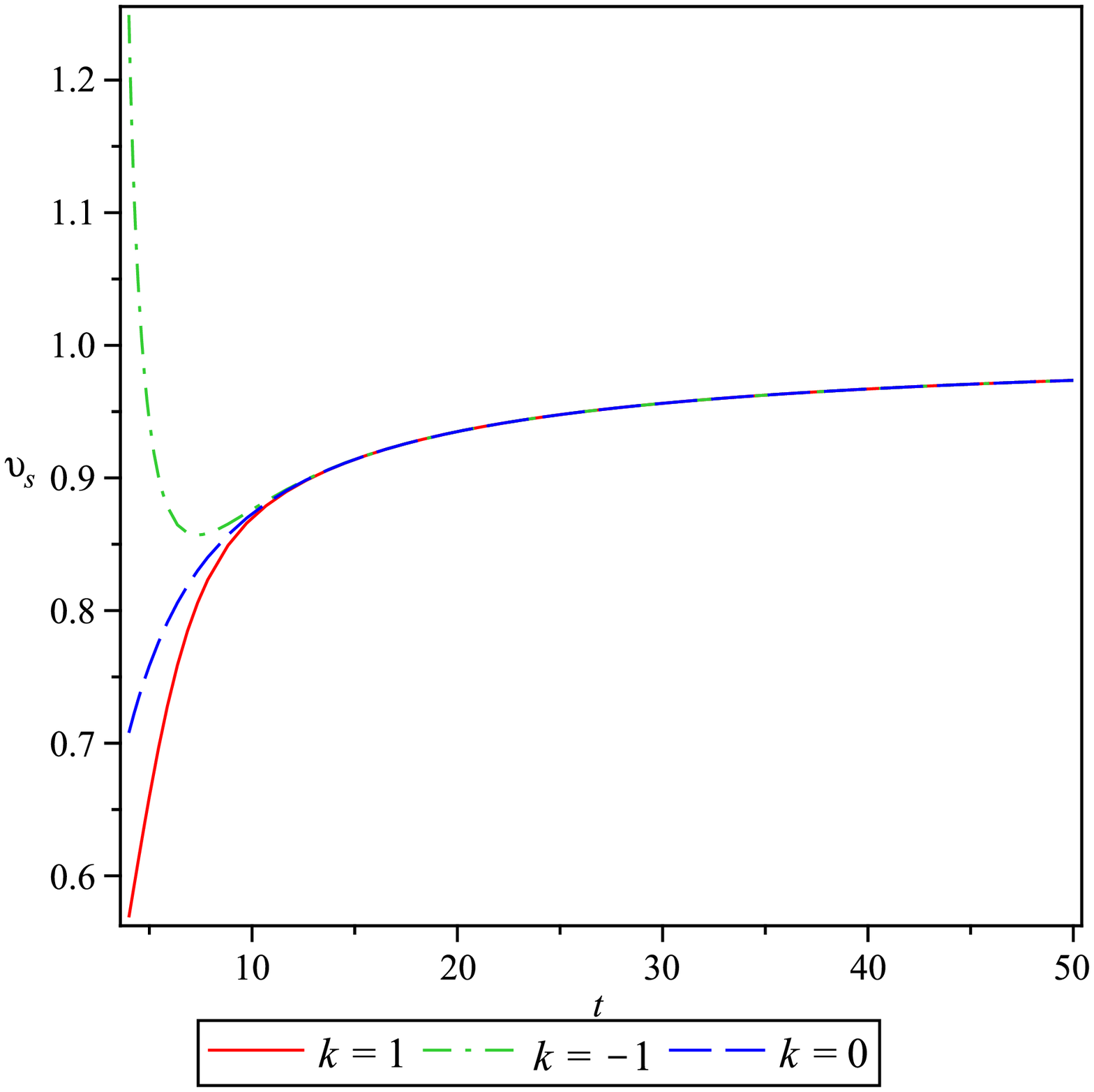}
\caption{The plot of sound speed $\upsilon_{s} $ Vs. $t$ for interacting two-fluid scenario}
\end{figure}
\begin{figure}[ht]
\centering
\includegraphics[width=8cm,height=8cm,angle=0]{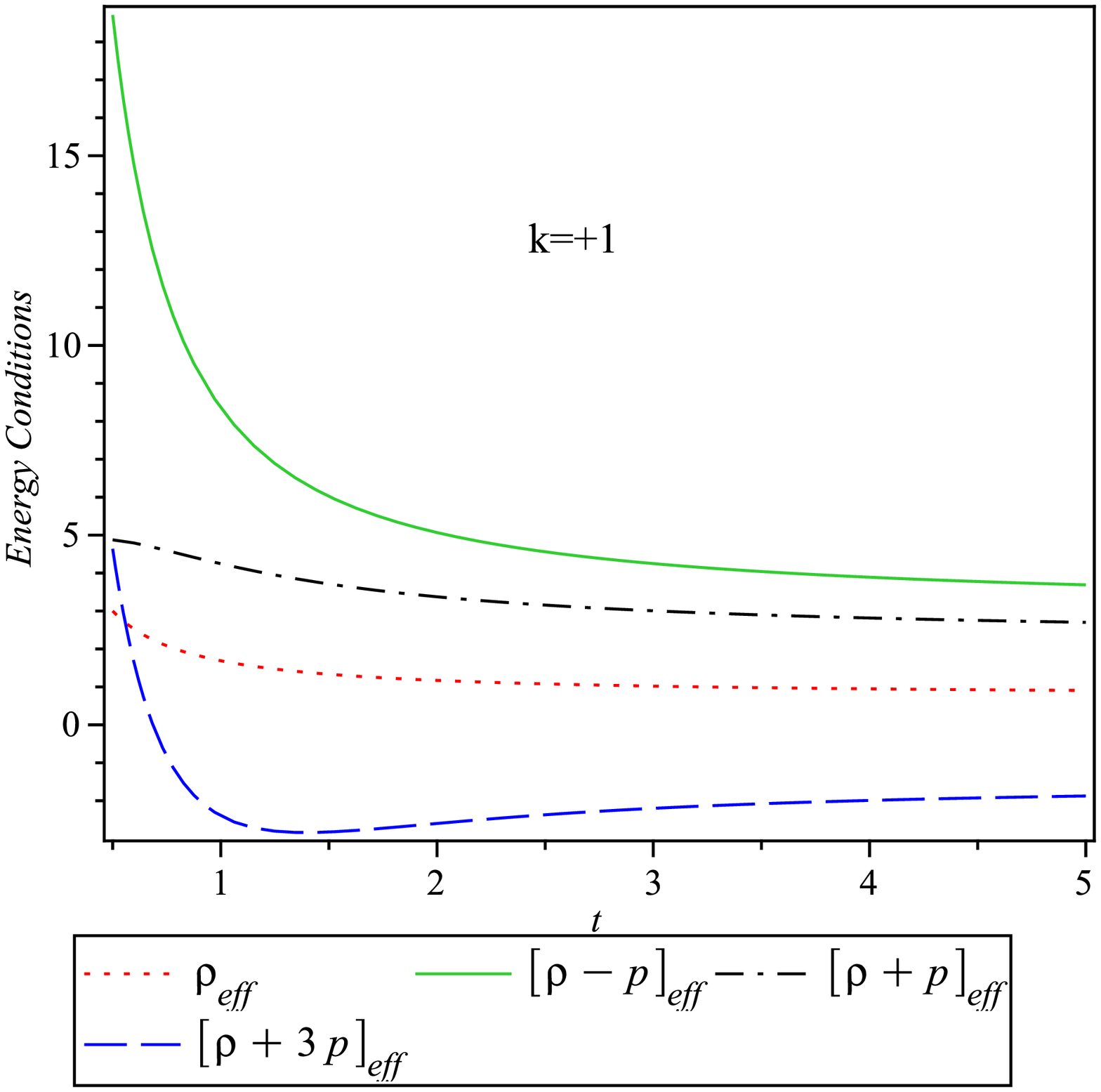} \\
\caption{The plot of energy conditions versus $t$ in non-interacting two-fluid model}
\end{figure}
\begin{figure}[ht]
\centering
\includegraphics[width=8cm,height=8cm,angle=0]{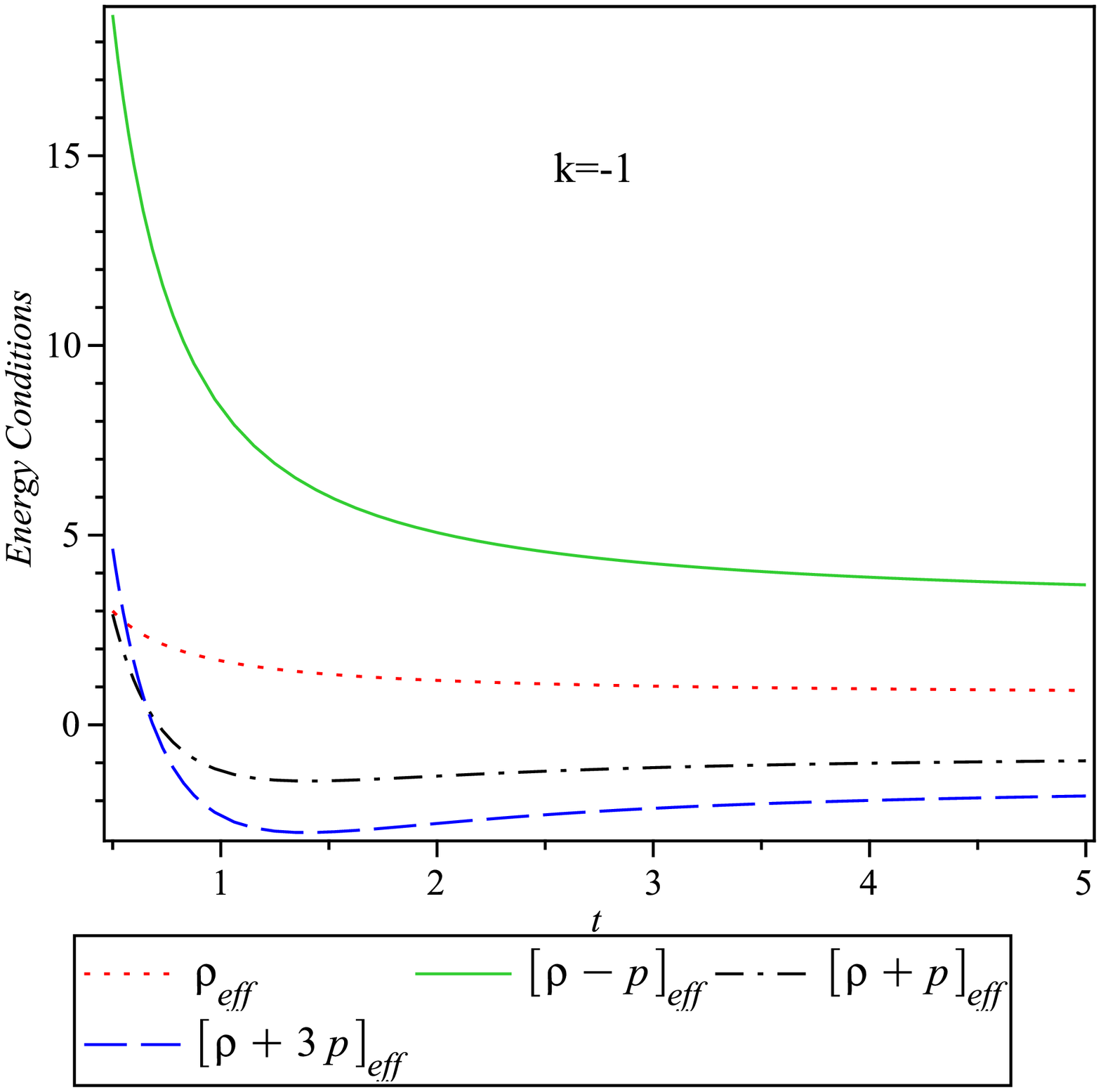} \\
\caption{The plot of energy conditions versus $t$ in non-interacting two-fluid model}
\end{figure}
\begin{figure}[ht]
\centering
\includegraphics[width=8cm,height=8cm,angle=0]{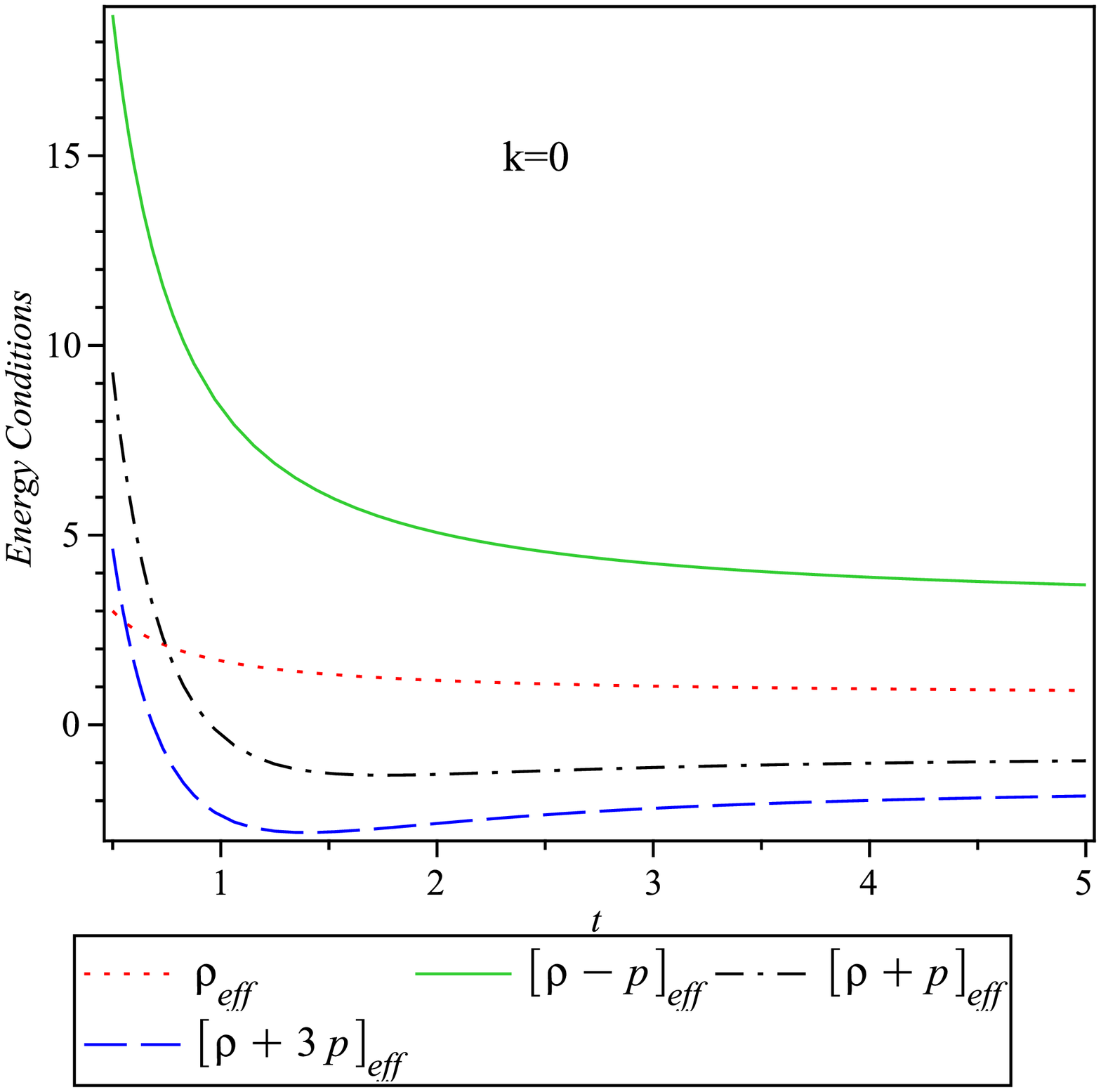} \\
\caption{The plot of energy conditions versus $t$ in non-interacting two-fluid model}
\end{figure}
\begin{figure}[ht]
\centering
\includegraphics[width=8cm,height=8cm,angle=0]{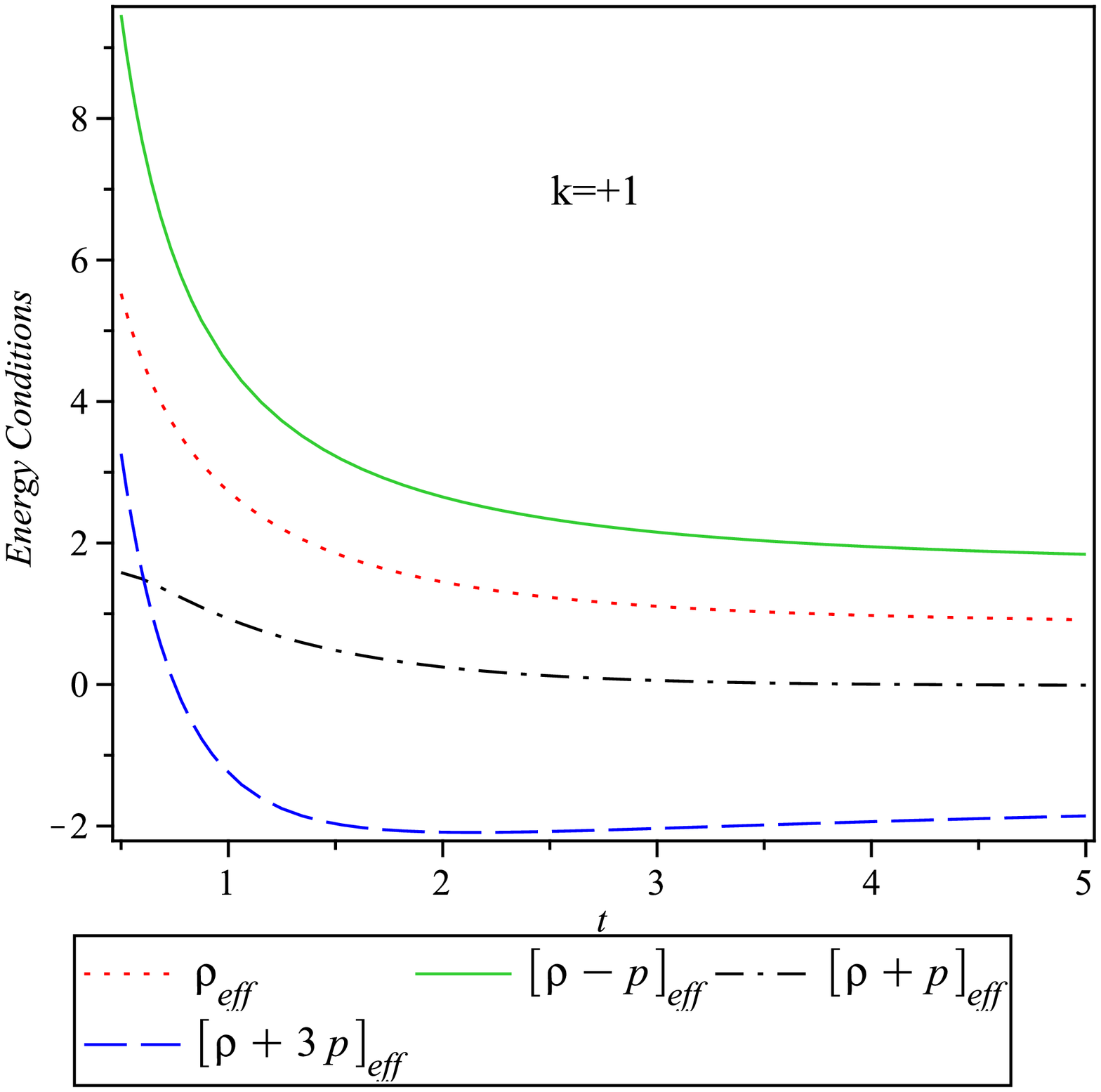} \\
\caption{The plot of energy conditions versus $t$ in interacting two-fluid model}
\end{figure}
\begin{figure}[ht]
\centering
\includegraphics[width=8cm,height=8cm,angle=0]{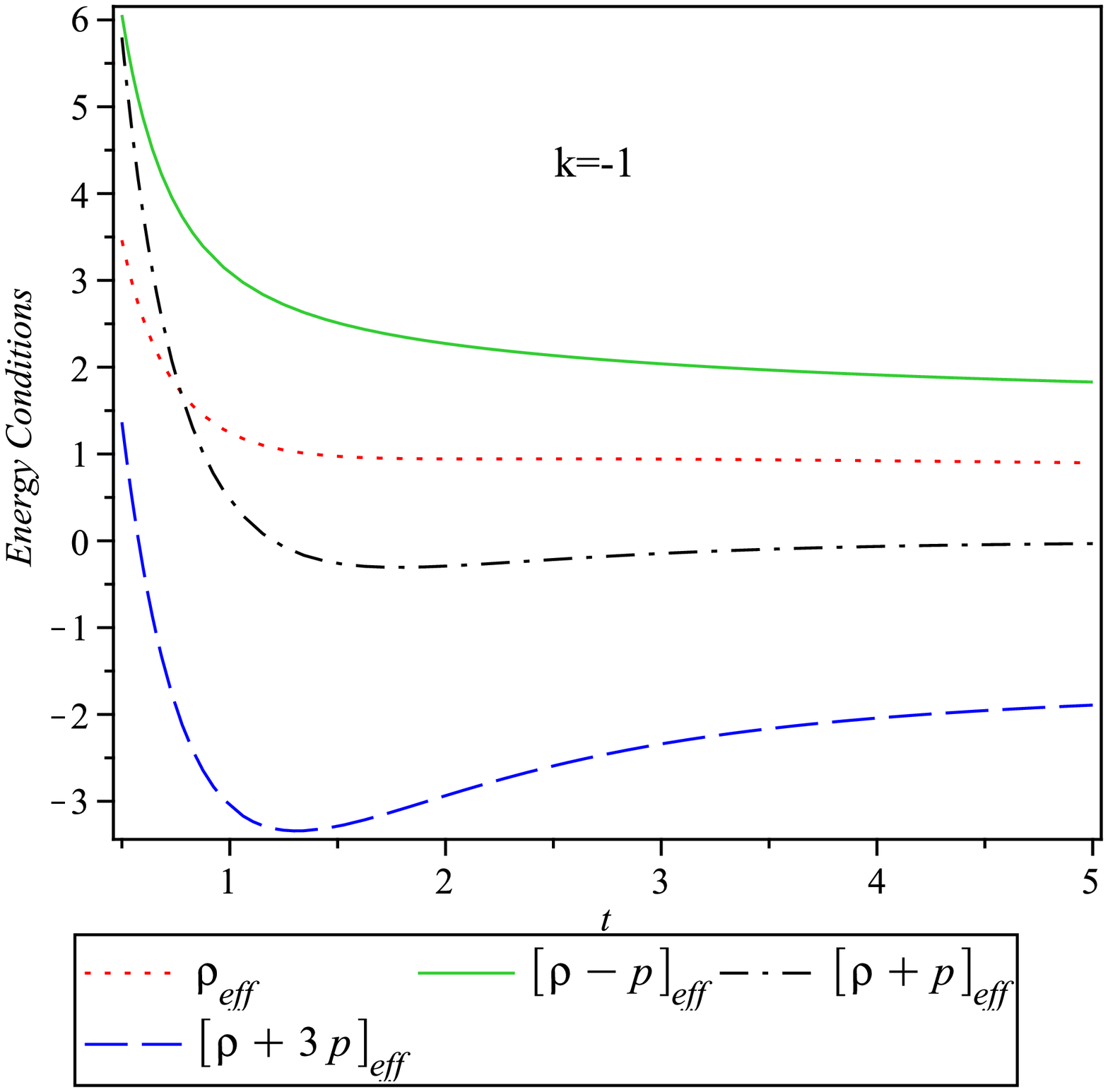} \\
\caption{The plot of energy conditions versus $t$ in interacting two-fluid model}
\end{figure}
\begin{figure}[ht]
\centering
\includegraphics[width=8cm,height=8cm,angle=0]{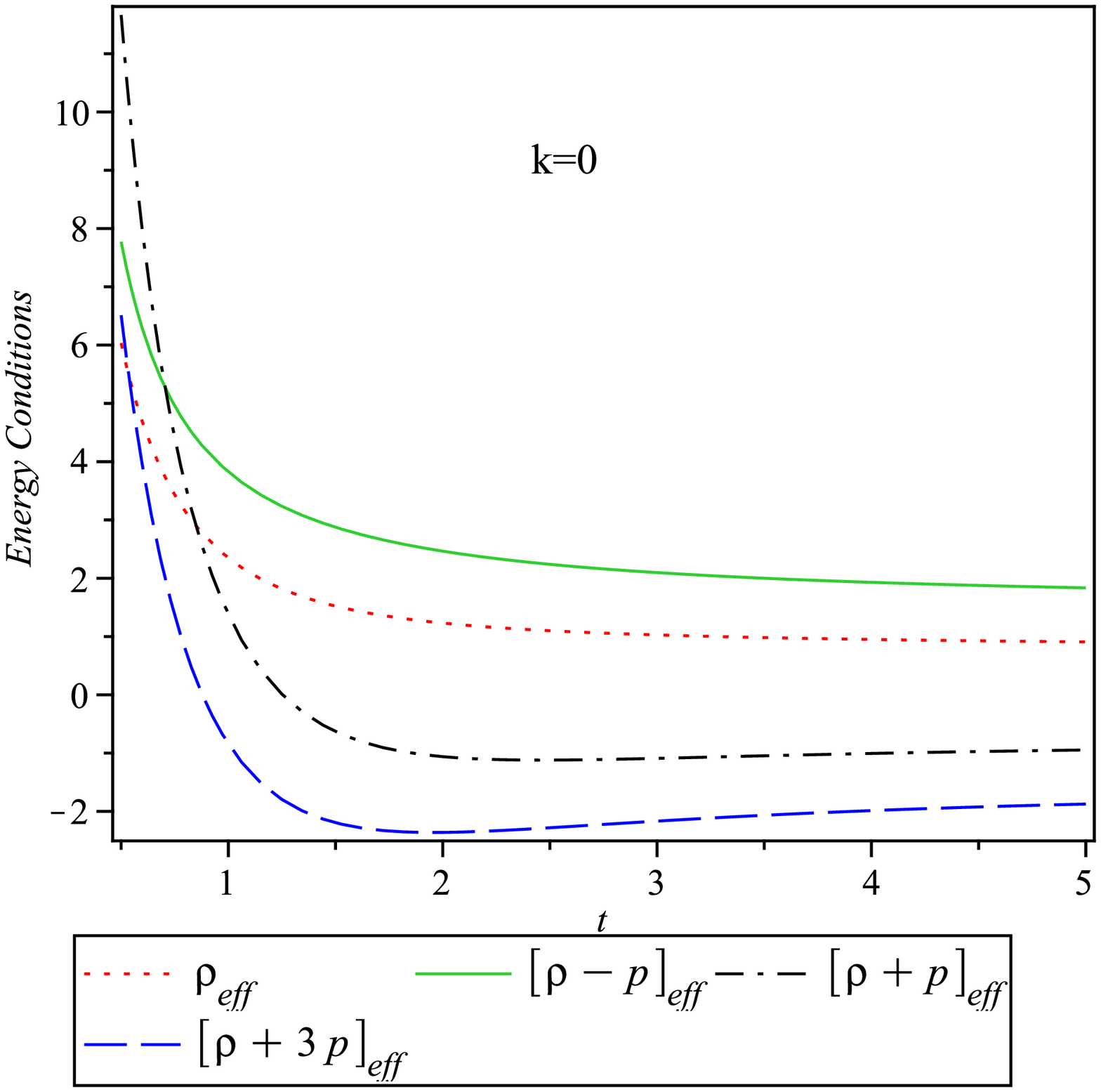} \\
\caption{The plot of energy conditions versus $t$ in interacting two-fluid model}
\end{figure}
\begin{figure}[ht]
\centering
\includegraphics[width=8cm,height=8cm,angle=0]{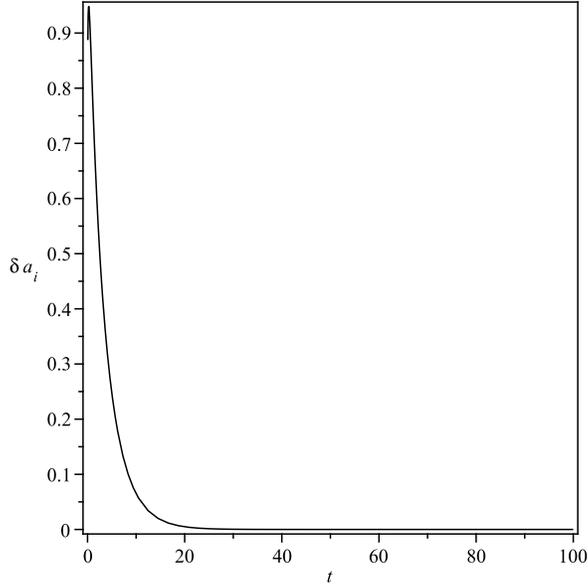} \\
\caption{The plot of the ``actual`` fluctuations $\delta a_{i}$ versus $t$}
\end{figure}

The expressions for the matter-energy density $\Omega_{m}$ and dark-energy density $\Omega_{D}$ are given by
\begin{equation}
\label{eq19}\Omega_{m} = \frac{\rho_{m}}{3H^{2}} = \frac{4\rho_{0}t^{2}}{3(t + n)^{2}}(t^{n}e^{t})^
{-\frac{3}{2}(1 + \omega_{m})},
\end{equation}
and
\begin{equation}
\label{eq20}\Omega_{D} = 1 + \frac{4k}{t^{n-2}e^{t}(n + t)^{2}} - \frac{4\rho_{0}t^{2}}{3(t + n)^{2}}(t^{n}
e^{t})^{-\frac{3}{2}(1 + \omega_{m})},
\end{equation}
respectively. Eqs. (\ref{eq19}) and (\ref{eq20}) reduce to
\begin{equation}
\label{eq21}\Omega = \Omega_{m} + \Omega_{D} = 1+\frac{4k}{t^{n-2}e^{t}(n+t)^{2}}.
\end{equation}
From the right hand side of Eq. (\ref{eq21}) it is clear that in flat universe ($k = 0$), $\Omega = 1$ and 
in open universe ($k = -1$), $\Omega < 1$ and in closed universe ($k = + 1$), $\Omega > 1$. But at late time 
we see for all flat, open and closed universes $\Omega \to 1$. This result is also compatible with the 
observational results. Since our model predicts a flat universe for large times and the present-day universe 
is very close to flat, so the derived model is also compatible with the observational results. The variation 
of density parameter with cosmic time has been shown in Fig. $2$. \\

In the Universe nearly $70\%$ of the energy is in the form of dark energy. Baryonic matter amounts to only 
$3-4\%$, while the rest of the matter ($27\%$ is believed to be in the form of a non-luminous component of 
non-baryonic nature with a dust-like equation of state ($w = 0$) known as cold dark matter (CDM). In this case, 
if the dark energy is composed just by a cosmological constant, then this scenario is called $\Lambda$-CDM model. 
A convenient method to describe models close to $\Lambda$ CDM is based on the cosmic jerk parameter $j$, a
dimensionless third derivative of the scale factor with respect to the cosmic time (Chiba and Nakamura 1998; Sahni 2002; 
Blandford et al. 2004; Visser 2004, 2005). A deceleration-to-acceleration transition occurs for models with a positive 
value of $j_{0}$ and negative $q_{0}$. Flat $\Lambda$ CDM models have a constant jerk $j = 1$. The jerk parameter 
in cosmology is defined as the dimensionless third derivative of the scale factor with respect to cosmic time
\begin{equation}
\label{eq22} j(t) = \frac{1}{H^{3}}\frac{\dot{\ddot{a}}}{a} \; ,
\end{equation}
and in terms of the scale factor to cosmic time
\begin{equation}
\label{eq23} j(t) = \frac{(a^{2}H^{2})^{''}}{2H^{2}} \; ,
\end{equation}
where the `dots' and `primes' denote derivatives with respect to cosmic time and scale factor, respectively.
The jerk parameter appears in the fourth term of a Taylor expansion of the scale factor around $a_{0}$
\begin{equation}
\label{eq24} \frac{a(t)}{a_{0}} = 1 + H_{0}(t - t_{0}) - \frac{1}{2}q_{0}H_{0}^{2}(t - t_{0})^{2} +
\frac{1}{6}j_{0}H_{0}^{3}(t - t_{0})^{3} + O\left[(t - t_{0})^{4}\right],
\end{equation}
where the subscript $0$ shows the present value. One can rewrite Eq. (\ref{eq22}) as
\begin{equation}
\label{eq25} j(t) = q + 2q^{2} - \frac{\dot{q}}{H}.
\end{equation}
Eqs. (\ref{eq15}) and (\ref{eq25}) reduce to
\begin{equation}
\label{eq27} j(t) = 1 - \frac{6n}{(n + t)^{2}} + \frac{8n}{(n + t)^{3}}.
\end{equation}
This value is overlap with the value $j\simeq2.16$ obtained from the combination of three kinematical data sets: the 
gold sample of type Ia supernovae (Riess et al. 2004), the SNIa data from the SNLS project (Astier et al. 2006), 
and the X-ray galaxy cluster distance measurements (Rapetti et al. 2007) for
\begin{equation}
\label{eq27} t = A - \frac{50n}{A} - n,
\end{equation}
where
\[
A = 0.03\left(84100n + 1450\sqrt{1450n^{3} + 3364n^{2}}\right)^{\frac{1}{3}}.
\]
\section{Interacting two-fluid model}
Secondly, we consider the interaction between dark energy and barotropic fluids. For this purpose we can write
The continuity equations for dark fluid and barotropic fluids as
\begin{equation}
\label{eq28} \dot{\rho}_{m} + 3\frac{\dot{a}}{a}(\rho_{m} + p_{m}) = Q,
\end{equation}
and
\begin{equation}
\label{eq29} \dot{\rho}_{D} + 3\frac{\dot{a}}{a}(\rho_{D} + p_{D}) = -Q.
\end{equation}
The quantity $Q$ expresses the interaction between the dark energy components. Since we are interested in an energy 
transfer from the dark energy to dark matter, we consider $Q > 0$. $Q > 0$, ensures that the second law of 
thermodynamics is fulfilled (Pavon and Wang 2009). Here we emphasize that the continuity Eqs. (\ref{eq28}) and
(\ref{eq29}) imply that the interaction term ($Q$) should be  proportional to a quantity with units of inverse of
time i.e $Q\propto \frac{1}{t}$. Therefore, a first and natural candidate can be the Hubble factor $H$ multiplied
with the energy density. Following Amendola et al. (2007) and Gou et al. (2007), we consider
\begin{equation}
\label{eq30} Q = 3H \sigma \rho_{m},
\end{equation}
where $\sigma$ is a coupling constant. Using Eq. (\ref{eq30}) in Eq. (\ref{eq28}) and after integrating, we 
obtain
\begin{equation}
\label{eq31} \rho_{m} = \rho_{0}a^{-3(1 + \omega_{m} - \sigma)}.
\end{equation}
By using Eq. (\ref{eq31}) in Eqs. (\ref{eq3}) and (\ref{eq4}), we again obtain the $\rho_{D}$ and $p_{D}$ 
in term of scale factor $a(t)$.
\begin{equation}
\label{eq32} \rho_{D} = 3\left(\frac{\dot{a}^{2}}{a^{2}} + \frac{k}{a^{2}}\right) - \rho_{0}a^{-3(1 + \omega_{m} - \sigma)},
\end{equation}
and
\begin{equation}
\label{eq33} p_{D} = -\left(2\frac{\ddot{a}}{a} + \frac{\dot{a}^{2}}{a^{2}} + \frac{k}{a^{2}}\right) - \rho_{0}(\omega_{m}
-\sigma)a^{-3(1 + \omega_{m} - \sigma)},
\end{equation}
respectively. Putting the value of $a(t)$ from Eq. (\ref{eq13}) in Eqs. (\ref{eq32}) and (\ref{eq33}), we obtain
\begin{equation}
\label{eq34} \rho_{D} = 3\left(\frac{n + t}{2t}\right)^{2} + \frac{3k}{t^{n}e^{t}} - \rho_{0}(t^{n}e^{t})^
{-\frac{3}{2}(1 + \omega_{m} - \sigma)},
\end{equation}
and
\begin{equation}
\label{eq35} p_{D} = -\left[3\left(\frac{n + t}{2t}\right)^{2} - \frac{n}{t^{2}} + \frac{k}{t^{n}e^{t}} - 
\rho_{0}(\omega_{m} - \sigma)(t^{n}e^{t})^{-\frac{3}{2}(1 + \omega_{m} - \sigma)}\right],
\end{equation}
respectively. Using Eqs. (\ref{eq34}) and (\ref{eq35}) in Eq. (\ref{eq7}), we can find the EoS parameter of dark
field as
\begin{equation}
\label{eq36} \omega_{D} = -\frac{3\left(\frac{n + t}{2t}\right)^{2} - \frac{n}{t^{2}} + \frac{k}{t^{n}e^{t}} 
-\rho_{0}(\omega_{m}-\sigma)(t^{n}e^{t})^{-\frac{3}{2}(1 + \omega_{m} - \sigma)}}{3\left(\frac{n + t}{2t}
\right)^{2} + \frac{3k}{t^{n}e^{t}} - \rho_{0}(t^{n}e^{t})^{-\frac{3}{2}(1 + \omega_{m} - \sigma)}}.
\end{equation}
The behavior of EoS for DE ($\omega_{D}$) in term of cosmic time $t$ is shown in Fig. $4$. It is observed that 
for closed universe the $\omega_{D}$ is decreasing function of time whereas for open and flat universes the EoS 
parameter is an increasing function of time, the rapidity of their growth at the early stage depends on the type 
the universes, while later on they all three tend to the same constant value independent to it. From this figure we also 
observe that the EoS parameters of closed, open and flat universes are varying in quintessence ($\omega_{D} > -1$), 
phantom ($-3 < \omega_{D} < -1$) and Super phantom ($\omega_{D} < -3$) regions respectively, while later on they 
tend to the same constant value $-1$ (i.e. cosmological constant) independent to it. From Fig. $4$, we also observe 
that the interaction pushes all closed, open and flat universes to darker regions. In this case we also observe that 
there is a ``little rip'' in the evolution of the open and flat universes. \\

The expressions for the matter-energy density $\Omega_{m}$ and dark-energy density $\Omega_{D}$ are given by
\begin{equation}
\label{eq37} \Omega_{m} = \frac{\rho_{m}}{3H^{2}} = \frac{4\rho_{0}t^{2}}{3(t + n)^{2}}(t^{n}e^{t})^
{-\frac{3}{2}(1 + \omega_{m} - \sigma)},
\end{equation}
and
\begin{equation}
\label{eq38} \Omega_{D} = 1 + \frac{4k}{t^{n-2}e^{t}(n + t)^{2}} - \frac{4\rho_{0}t^{2}}{3(t + n)^{2}}
(t^{n}e^{t})^{-\frac{3}{2}(1 + \omega_{m} - \sigma)},
\end{equation}
respectively. From Eqs. (\ref{eq37}) and (\ref{eq38}), we obtain
\begin{equation}
\label{eq39}\Omega = \Omega_{m} + \Omega_{D} = 1 + \frac{4k}{t^{n-2}e^{t}(n + t)^{2}}.
\end{equation}
which is the same as Eq. (\ref{eq21}). Therefore, we observe that in interacting case the density parameter 
has the same properties an non-interacting case. The expressions for deceleration parameter and jerk parameter 
are also same as in the case of non-interacting case.\\

Studying the interaction between the dark energy and ordinary matter will open a possibility of detecting the 
dark energy. It should be pointed out that evidence was recently provided by the Abell Cluster A586 in support 
of the interaction between dark energy and dark matter (Bertolami et al. 2007; Delliou et al. 2007). We observe 
that in non-interacting case both open and flat universes can cross the phantom region whereas in interacting 
case only open universe can cross phantom region.

\section{Physical acceptability and stability of solutions}
For the stability of corresponding solutions in both non-interacting and interacting models, we should check 
that our models are physically acceptable. For this, firstly it is required that the velocity of sound should be less 
than velocity of light i.e. within the range $0 \leq \upsilon_{s} = \left(\frac{dp}{d\rho}\right) \leq 1$. \\

In our non-interacting and interacting models, we obtained the sound speeds as 
\begin{equation}
\label{eq40}
\upsilon_{s} = \frac{-\frac{3}{2}(\frac{0.5 + t}{t^{2}}) + \frac{3}{2}\frac{(0.5 + t)^{2}}{t^{3}} - \frac{1}{t^{3}} + 
\frac{1}{2}\frac{k}{t^{\frac{3}{2}}e^{t}} +\frac{k}{\sqrt{t}e^{t}} - \frac{1.125(\frac{1}{2}\frac{e^{t}}{\sqrt{t}} + 
\sqrt{t}e^{t})}{(\sqrt{t}e^{t})^\frac{13}{4}}}{\frac{3}{2}(\frac{0.5 + t}{t^{2}}) - \frac{3}{2}\frac{(0.5 + t)^{2}}{t^{3}} 
- \frac{3}{2}\frac{k}{t^{\frac{3}{2}}e^{t}} - \frac{3k}{\sqrt{t}e^{t}} +\frac{9}{4}\frac{\frac{1}{2}\frac{e^{t}}
{\sqrt{t}} + \sqrt{t}e^{t}}{(\sqrt{t}e^{t})^\frac{13}{4}}},
\end{equation}
and 
\begin{equation}
\label{eq41}
\upsilon_{s} = \frac{-\frac{3}{2}(\frac{0.5 + t}{t^{2}}) + \frac{3}{2}\frac{(0.5 + t)^{2}}{t^{3}} - \frac{1}{t^{3}} + 
\frac{1}{2}\frac{k}{t^{\frac{3}{2}}e^{t}} + \frac{k}{\sqrt{t}e^{t}} - \frac{0.36(\frac{1}{2}\frac{e^{t}}{\sqrt{t}} + 
\sqrt{t}e^{t})}{(\sqrt{t}e^{t})^{2.8}}}{\frac{3}{2}(\frac{0.5 + t}{t^{2}}) - \frac{3}{2}\frac{(0.5 + t)^{2}}{t^{3}} - 
\frac{3}{2}\frac{k}{t^{\frac{3}{2}}e^{t}} - \frac{3k}{\sqrt{t}e^{t}} + \frac{9}{4}\frac{\frac{1}{2}\frac{e^{t}}{\sqrt{t}} 
+ \sqrt{t}e^{t}}{(\sqrt{t}e^{t})^{2.8}}},
\end{equation}
respectively. In both cases we observe that $\upsilon_{s} < 1$. From Figures $5$ \& $6$, we observe that in both non-
interacting and interacting cases  $\upsilon_{s} < 1$. \\

Secondly, the weak energy conditions (WEC) and dominant energy conditions (DEC) are given by \\

(i) $\rho_{eff} \geq 0$, ~ ~ (ii) $\rho_{eff} - p_{eff} \geq 0$ ~ ~ and ~ ~ (iii) $\rho_{eff} + p_{eff} \geq 0$. \\

The strong energy conditions (SEC) are given by $\rho_{eff} + 3p_{eff} \geq 0$. \\

From the Figures $7$ $-$ $12$, we observe that 
\begin{itemize}
\item The WEC and DEC for the closed universe in both non-interacting and interacting cases are satisfied.  
\end{itemize}
\begin{itemize}
\item In both open \& flat models, the WEC and DEC are satisfied in initial stages of the evolution of the universe (i.e.
in decelerating phase). In these both models WEC and SEC are violated at later times as expected but DEC does not violet.
\end{itemize}
\begin{itemize}
\item The SEC for both non-interacting and interacting cases are satisfied in early stages of the evolution of the universe 
whereas it violet at present epoch due to acceleration for all three open, closed and flat models as expected.
\end{itemize}
Therefore, on the basis of above discussions and analysis, our corresponding solutions are physically acceptable. \\ 

A rigorous analysis on the stability of the corresponding solutions can be done by invoking a perturbative approach.
Perturbations of the fields of a gravitational system against the background evolutionary solution should be checked to 
ensure the stability of the exact or approximated background solution (Chen and Kao 2001). Now we will study the stability 
of the background solution with respect to perturbations of the metric. Perturbations will be considered for all three 
expansion factors $a_{i}$ via
\begin{equation}
\label{eq42}
a_{i}\rightarrow a_{Bi}+\delta a_{i}=a_{Bi}(1+\delta b_{i})
\end{equation}
We will focus on the variables $\delta b_{i}$ instead of $\delta a_{i}$ from now on for convenience. Therefore, the 
perturbations of the volume scale factor $V_{B} = \Pi_{i=1}^{3}a_{i}$, directional Hubble factors $\theta_{i} = 
\frac{\dot{a_{i}}}{a_{i}}$ and the mean Hubble factor $\theta = \sum_{i=3}^{3}\frac{\theta_{i}}{3} = 
\frac{\dot{V}}{3V}$ can be shown to be
\begin{equation}
\label{eq43}
V\rightarrow V_{B} + V_{B}\sum_{i}\delta b_{i}, ~~~~ \theta_{i}\rightarrow \theta_{Bi}+\sum_{i}\delta 
b_{i}, ~~~~\theta\rightarrow \theta_{B}+\frac{1}{3}\sum_{i}\delta b_{i}
\end{equation}
One can show that the metric perturbations $\delta b_{i}$, to the linear order in $\delta b_{i}$, obey the following equations
\begin{equation}
\label{eq44}
\sum_{i}\delta\ddot{b_{i}} + 2\sum\theta_{Bi}\delta\dot{b_{i}} = 0,
\end{equation}
\begin{equation}
\label{eq45}
\delta\ddot{b_{i}} + \frac{\dot{V}_{B}}{V_{B}}\delta \dot{b_{i}} + \sum_{j}\delta\dot{b_{j}}\theta_{Bi} = 0,
\end{equation}
\begin{equation}
\label{eq46}
\sum\delta\dot{b_{i}} = 0.
\end{equation}
From above three equations, we can easily find
\begin{equation}
\label{eq47}
\delta \ddot{b_{i}}+\frac{\dot{V}_{B}}{V_{B}}\delta \dot{b_{i}}=0,
\end{equation}
where $V_{B}$ is the background volume scale factor. In our case, $V_{B}$ is given by
\begin{equation}
\label{eq48}
V_{B}=t^{\frac{3}{2}n}e^{\frac{3}{2}t}.
\end{equation}
Using above equation in equation (47) and after integration we get
\begin{equation}
\label{eq49}
\delta b_{i} = c_{i} t^{-\frac{3}{4}n}e^{-\frac{3}{4}t}\mbox{WittakerM}\left(-\frac{3}{4}n,-\frac{3}{4}n + 
\frac{1}{2},\frac{3}{2}t\right),
\end{equation} 
where $c_{i}$ is an integration constant. Therefore, the ``actual`` fluctuations for each expansion factor 
$\delta a_{i} = a_{Bi}\delta b_{i}$
is given by
\begin{equation}
\label{eq50}
\delta a_{i}\rightarrow c_{i} t^{-\frac{n}{4}}e^{-\frac{t}{4}}\mbox{WittakerM}\left(-\frac{3}{4}n,-\frac{3}{4}n + 
\frac{1}{2},\frac{3}{2}t\right).
\end{equation} 
From above equation we see that for $n> > 1$, $\delta a_{i}$ approaches zero. Fig. $13$ is the plot of the 
``actual`` fluctuations $\delta a_{i}$ versus $t$ which also shows that $\delta a_{i} \to 0$ as $t \to \infty$. 
Consequently, the background solution is stable against the perturbation of the graviton field. 

\section{Concluding remarks}
In this present work we continue and extend the previous work of Amirhashchi et al. (2011a).
In summary, we have studied a system of two fluid within the scope of a spatially homogeneous and isotropic 
FRW model. The role of two fluid either minimally or directly coupled in the evolution of the dark energy 
parameter has been investigated. The scale factor is considered to be a power law function of time which yields 
a time dependent deceleration parameter. It is observed that in an interacting and non-interacting cases both  
open and flat universes can cross the phantom region. It is observed that the closed universe is corresponding 
to quintessence whereas the flat and open universes are corresponding to phantom model of universe. During the 
evolution of the universe, we find that the EoS parameter for closed universe changes from $w > -1$ to $w < -1$, 
which is consistent with recent observations. If we put $n = 1$ in the present paper, we obtain all results of 
recent paper of Amirhashchi et al. (2011a). \\

It is observed that the EoS parameters $\omega_{D}$ of closed, open and flat universes are varying in quintessence 
($\omega_{D} > -1$), phantom ($-3 < \omega_{D} < -1$) and Super phantom ($\omega_{D} < -3$) regions respectively, 
while later on they tend to the same constant value $-1$ (i.e. cosmological constant region) independent to it. \\

Our special choice of scale factor yields a time dependent deceleration parameter which represents a model of 
Universe which takes evolution from decelerating to accelerating phase which is in good agreement with current 
observations. It is worth mentioned here that for different choice of $n$, we can generate a class of DE models 
in FRW universe. It is also observed that such DE models are also in good harmony with current observations. \\

We also observe that our corresponding solutions are physically acceptable and the solutions are stable. Thus, the 
solutions demonstrated in this paper may be useful for better understanding of the characteristic of DE in the 
evolution of universe within the framework of FRW. 
\section*{Acknowledgments}
One of the authors (A. Pradhan) would like to thank the Institute of Mathematical Sciences (IMSc.), Chennai 
(Madras), India for providing facility and support under associateship scheme where part of this work was 
carried out. The financial support (Project No. C.S.T./D-1536) in part by State Council of Science \& Technology, 
U.P., India is also gratefully acknowledged by A. Pradhan. Sincere thanks are due to the anonymous reviewer for 
his constructive comments to enhance the quality of the paper. 


\begin{thebibliography}{000}
\bibitem {ref1}
Abazajian, K., et al. [SDSS Collaboration]: Astron. J. {\bf 128}, 502 (2004). arXiv:astro-ph/0403325. 
\bibitem {ref2}
Akarsu, $\ddot{O}$., Kilinc, C.B.: Gen. Relat. Gravit. {\bf 42}, 119 (2010a)
\bibitem {ref3}
Akarsu, $\ddot{O}$., Kilinc, C.B.: Gen. Relat. Gravit. {\bf 42}, 763 (2010b)
\bibitem {ref4}
Alimohammadi, M., Mohseni Sadjadi, H.: Phys. Rev. D {\bf 73}, 083527 (2006)
\bibitem {ref5}
Amanullah, R., et al.: Ap. J. {\bf 716}, 712 (2010)
\bibitem {ref6}
Amendola, L.: Mon. Not. R. Astron. Soc. {\bf 342}, 221 (2003)
\bibitem {ref7}
Amendola, L., Camargo Campos, G., Rosenfeld, R.: Phys. Rev. D {\bf 75}, 083506 (2007)
\bibitem{ref8}
Amirhashchi, H., Pradhan, A., Saha, B.: Chin. Phys. Lett. {\bf 28}, 039801 (2011a)
\bibitem {ref9}
Amirhashchi, H., Pradhan, A., Saha, B.: Astrophys. Space Sci. {\bf 333}, 295 (2011c)
\bibitem {ref10}
Amirhashchi, H., Pradhan, A., Zainuddin, H.: Int. J. Theor. Phys. {\bf 50}, 3529 (2011b)
\bibitem {ref11}
Astashenok, A.V., Nojiri, S., Odintsov, S.D., Yurov, A.V.: (2012). arXiv:1201.4056[gr-qc]
\bibitem {ref12}
Astier, P., et al.: Astron. Astrophys. {\bf 447}, 31 (2006)
\bibitem {ref13}
Bennett, C.L., et al.: Astrophys. J. Suppl. {\bf 148}, 1 (2003)
\bibitem {ref14}
Bertolami, O., Gil Pedro, F., Le Delliou, M.: Phys. Lett. B {\bf 654}, 165 (2007)
\bibitem {ref15}
Blandford, R.D., Amin, M., Baltz, E.A., Mandel, K., Marshall, P.J.: (2004). arXiv:astro-ph/0408279
\bibitem {ref16}
Brevik, I., Elizalde, E., Nojiri, S., Odintsov, S.D.: arXiv:1107.4642[hep-th] 2011
\bibitem {ref17}
Cai, Y., Li, M., Lu, J., Piao, Y., Qiu, T., Zhang, X.: Phys. Lett. B {\bf 651}, 1 (2007)
\bibitem {ref18}
Caldwell, R.R., Dave, R., Steinhardt, P.J.: Phys. Rev. Lett. {\bf 80}, 1582 (1998)
\bibitem{ref19}
Caldwell, R.R.: Phys. Lett. B {\bf 545}, 23 (2002)
\bibitem {ref20}
Chen, C.-Mei, Kao, W.F.: (2011). arXiv:hep-th/0104101
\bibitem {ref21}
Chiba, T., Nakamura, T.: Prog. Theor. Phys. {\bf 100}, 1077 (1998)
\bibitem {ref22}
Chimento, L.P., Jakubi, A.S., Pavon, D., Zimdahl, W.: Phys. Rev. D {\bf 67}, 083513 (2003)
\bibitem {ref23}
Chimento, L.P., Pavon, D.: Phys. Rev. D {\bf 73}, 063511 (2006)
\bibitem {ref24}
Clocchiatti, A., et al.: Astrophys. J. {\bf 642}, 1 (2006)
\bibitem {ref25}
Copeland, E.J., Sami, M., Tsujikawa, S.: Int. J. Mod. Phys. D {\bf 15}, 1753 (2006)
\bibitem {ref26}
de Bernardis, P.: et al.: Nature {\bf 391}, 5 (1998)
\bibitem{ref27}
Elizalde, E., Nojiri, S., Odintsov, S.D.: Phys. Rev. D {\bf 70}, 043539 (2004)
\bibitem {ref28}
Feng, B., Wang, X., Zhang, X.: Phys. Lett. B {\bf 607}, 35 (2005)
\bibitem {ref29}
Frampton, P.H., Ludwick, K.J., Scherrer, R.J.: Phys. Rev. D {\bf 84}, 063003 (2011). arXiv:1106.4996[astro-ph.CO]
\bibitem {ref30}
Frampton, P.H., Ludwick, K.J., Nojiri, S., Odintsov, S.D., Scherrer, R.J.: 92012a). arXiv:1108.0067[hep-th]
\bibitem {ref31}
Frampton, P.H., Ludwick, K.J., Scherrer, R.J.: Phys. Rev. D {\bf 85}, 083001 (2012b). arXiv:1108.0067[hep-th]
\bibitem {ref32}
Frampton, P.H., Takahashi, T.: Phys. Lett. B {\bf 557}, 135 (2003). arXiv:astro-ph/0211544
\bibitem {ref33}
Guo, Z.K., Ohta, N., Tsujikawa, S.: Phys. Rev. D {\bf 76}, 023508 (2007)
\bibitem{ref34}
Granda, L.N., Loaiza, E.: Int. Jour. Mod. Phys. D {\bf 21}, 1250002 (2012). arXiv:1111.2454[hep-th] 
\bibitem{ref35}
Hanany, S., et al.: Astrophys. J. {\bf 493}, L53 (2000)
\bibitem {ref36}
Hawkins, E., et al.: Mon. Not. Roy. Astron. Soc. {\bf 346}, 78 (2003). arXiv:astro-ph/0212375
\bibitem {ref37}
Kowalski, M., et al.: Astrophys. J. {\bf 686}, 749 (2008)
\bibitem {ref38}
Kumar, S., Yadav, A.K.: Mod. Phys. Lett. A {\bf 26}, 647 (2011)
\bibitem {ref39}
Kumar, S., Singh, C.P.: Gen. Rel. Grav. {\bf 43}, 1427 (2011)
\bibitem {ref40}
Le Delliou, M., Bertolami, O., Gil Pedro, F.: AIP Conf. Proc. {\bf 957}, 421 (2007)
\bibitem{ref41}
Liang, N.M., Gao, C.J., Zhang, S.N.: Chin. Phys. Lett. {\bf 26}, 069501 (2009)
\bibitem {ref42}
Nesseris, S., Perivolaropoulos, L.: Phys. Rev. D {\bf 70}, 123529 (2004). arXiv:astro-ph/0410309
\bibitem {ref43}
Nojiri, S., Odintsov, S.D.: Phys. Lett. B {\bf 571}, 1 (2003)
\bibitem {ref44}
Nojiri, S., Odintsov, S.D., Saez-Gomez, D: 2012. arXiv:1108.0767[hep-th] 
\bibitem {ref45}
Nojiri, S., Odintsov, S.D., Tsujikawa, S.: Phys. Rev. D {\bf 71}, 063004 (2005)
\bibitem {ref46}
Padmanabhan, T., Roychowdhury, T.: Mon. Not. R. Astron. Soc. {\bf 344}, (2003) 823
\bibitem {ref47}
Padmanabhan, T.: Phys. Rep. {\bf 380}, 235 (2003)
\bibitem {ref48}
Pavon, D., Wang, B.: Gen. Relativ. Gravit. {\bf 41}, 1 (2009)
\bibitem {ref49}
Peebles, P.J.E., Ratra, B.: Rev. Mod. Phys. {\bf 75}, 559 (2003)
\bibitem {ref50}
Perlmutter, S., et al. [Supernova Cosmology Project Collaboration]: Astrophys. J. {\bf 517}, 565 (1999).
arXiv:astro-ph/9812133
\bibitem {ref51}
Pradhan, A., Amirhashchi, H., Saha, B: Astrophys. Space Sci. {\bf 333}, 343 (2011a)
\bibitem {ref52}
Pradhan, A., Amirhashchi, H., Saha, B: Int. J. Theor. Phys. {\bf 50}, 2923 (2011b)
\bibitem {ref53}
Pradhan, A., Amirhashchi, H.: Mod. Phys. Lett. A {\bf 26}, 2261 (2011a)
\bibitem {ref54}
Pradhan, A., Amirhashchi, H.: Astrophys. Space Sci. {\bf 332}, 441 (2011b)
\bibitem {ref55}
Pradhan, A., Jaiswal, R., Jotania, K., Khare, R.K.: Astrophys. Space Sci. {\bf 337}, 401 (2012).
\bibitem {ref56}
Rapetti, D., Allen, S.W., Amin, M.A., Blandford, R.D.: Mon. Not. Roy. Astron. Soc. {\bf 375}, 1510 (2007)
\bibitem {ref57}
Ratra, B., Peebles, P.E.J.: Phys. Rev. D {\bf 37}, 3406 (1988)
\bibitem {ref58}
Riess, A.G., et al. [Supernova Search Team Collaboration]: Astron. J. {\bf 116}, 1009 (1998). arXiv:astro-ph/9805201
\bibitem {ref59}
Riess, A.G., et al.: Astrophys. J. {\bf 560}, 49 (2001)
\bibitem {ref60}
Riess, A.G., et al.: Astrophys. J. {\bf 607}, 665  (2004)
\bibitem {ref61}
Sahni, V., Starobinsky, A.: Int. J. Mod. Phys. D {\bf 9}, 273 (2000)
\bibitem {ref62}
Sahni, V., Shtanov, Y.: J. Cosmol. Astropart. Phys. {\bf 11}, 014 (2003) 
\bibitem {ref63}
Sahni, V.: (2002). arXiv:astro-ph/0211084
\bibitem {ref64}
Setare, M.R.: Phys. Lett. B {\bf 641}, 130 (2006)
\bibitem {ref65}
Setare, M.R., Sadeghi, J., Amani, A.R.: Phys. Lett. B {\bf 673}, 241 (2009)
\bibitem {ref66}
Spergel, D.N., et al. [WMAP Collaboration]: Astrophys. J. Suppl. {\bf 148}, 175 (2003). arXiv:astro-ph/0302209
\bibitem {ref67}
Tegmark, M., et al. [SDSS Collaboration]: Phys. Rev. D {\bf 69}, 103501 (2004). arXiv:astro-ph/0310723
\bibitem {ref68}
Tonry, J.L., et al.: Astrophys. J. {\bf 594}, 1 (2003)
\bibitem {ref69}
Verde, L., et al.: Mon. Not. Roy. Astron. Soc. {\bf 335}, 432 (2002). arXiv:astro-ph/0112161 
\bibitem {ref70}
Visser, M: Class. Quantum Grav. {\bf 21}, 2603 (2004)
\bibitem {ref71}
Visser, M.: Gen. Relativ. Gravit. {\bf 37}, 1541 (2005)
\bibitem {ref72}
Wei, H., Tang, N.N., Zhang, S.N.: Phys. Rev. D {\bf 75}, 043009 (2007)
\bibitem {ref73}
Weinberg, S.: Rev. Mod. Phys. {\bf 61}, 1 (1989)
\bibitem {ref74}
Wetterich, C.: Phys. Lett B {\bf 302}, 668 (1988)
\bibitem {ref75}
Xi, P., Zhai, X., Li, X.: Phys. Lett. B {\bf 706}, 482 (2012)
\bibitem {ref76}
Xiangyun, Fu., Hongwei, Yu., Puxun, Wu.: Phys. Rev. D {\bf 78}, 063001 (2008)
\bibitem {ref77}
Yadav, A.K.: Astrophys. Space Sci. {\bf 335}, 565 (2011)
\end{thebibliography}
\end{document}